\documentclass[aps,prl,preprint,groupedaddress]{revtex4-1}
\usepackage{times}
\usepackage{bm}
\usepackage[dvipdf]{graphicx}
\usepackage{amsmath}
\usepackage{bm}
\usepackage{natbib}
\usepackage{color}
\usepackage{epstopdf}
\usepackage{natbib}
\begin{document}
\title{Pseudospin-driven spin relaxation mechanism in graphene}

\author{Dinh Van Tuan$^{1,2}$, Frank Ortmann$^{1,3,4}$, David Soriano$^{1}$, Sergio O. Valenzuela$^{1,5}$ and Stephan Roche$^{1,5}$}

\affiliation{$^1$ICN2 - Institut Catala de Nanociencia i Nanotecnologia,
Campus UAB, 08193 Bellaterra (Barcelona), Spain\\
$^2$Department of Physics, Universitat Aut\'{o}noma de Barcelona, Campus UAB, 08193 Bellaterra, Spain\\
$^3$ Institute for Materials Science and Max Bergmann Center of Biomaterials, Technische Universit\"{a}t Dresden, 01062 Dresden, Germany\\
$^4$ Dresden Center for Computational Materials Science, TU Dresden, 01062 Dresden, Germany\\
$^5$ICREA, Instituci\'{o} Catalana de Recerca i Estudis Avan\c{c}ats,
 08070 Barcelona, Spain}

\maketitle

{\bf The possibility of transporting spin information over long distances in graphene, owing to its small intrinsic spin-orbit coupling (SOC) and the absence of hyperfine interaction, has led to intense research into spintronic applications. However, measured spin relaxation times are orders of magnitude smaller than initially predicted, while the main physical process for spin dephasing and its charge-density and disorder dependences remain unconvincingly described by conventional mechanisms. Here, we unravel a spin relaxation mechanism for nonmagnetic samples that follows from an entanglement between spin and pseudospin driven by random SOC, which makes it unique to graphene. The mixing between spin and pseudospin-related Berry's phases results in fast spin dephasing even when approaching the ballistic limit, with increasing relaxation times away from the Dirac point, as observed experimentally. The SOC can be caused by adatoms, ripples or even the substrate, suggesting novel spin manipulation strategies based on the pseudospin degree of freedom.}


The electronic properties of monolayer graphene strongly differ from those of two-dimensional metals and semiconductors in part because of inherent electron-hole band structure symmetry and a particular density of states which vanishes at the Dirac point \cite{CastroNeto2009}. Additionally, the sublattice degeneracy and honeycomb symmetry lead to eigenstates that hold an additional quantum (Berry's) phase, associated with the so-called pseudospin quantum degree of freedom. All of these electronic features are manifested through the Klein tunneling phenomenon \cite{Katsnelson2006}, weak antilocalization \cite{McCann2006} or the anomalous quantum Hall effect \cite{Novoselov2007}. The possibility of using the pseudospin as a means to transport and store information has also been theoretically proposed \cite{Rycerz2007, Sanjose2009}. There, the role of the pseudospin is equivalent to that of the spin in spintronics, such as in the pseudospin analogue of the giant magnetoresistance in bilayer graphene \cite{Sanjose2009}.

Even though pseudospin-related effects drive most of the unique transport signatures of graphene, the role of the pseudospin on the spin relaxation mechanism has not been explicitly addressed and quantified. Pseudospin and spin dynamics are usually perceived as decoupled from each other, with pseudospin lifetimes being much shorter and pseudospin dynamics much faster than those for spins. However, this picture breaks down in the vicinity of the Dirac point, a region that is usually out of reach of perturbative approaches and that is particularly relevant for experiments, because Fermi energies can only be shifted by about 0.3 eV via electrostatic gating. Moreover, in the presence of SOC, spin couples to orbital motion, and therefore to pseudospin \cite{Rashba2009}, so that spin and pseudospin dynamics should not be treated independently.  Actually, any initially spin polarized state injected at low energy (for instance  $|\Psi\rangle\otimes | \uparrow\rangle$), should evolve under the time-evolution operator (including the spin-orbit coupling term) towards an entangled state $ \sim \left ( \begin{array}{c} 0 \\ 1 \end{array} \right ) \otimes | \uparrow \rangle \pm  \left ( \begin{array}{c} i\\ 0 \end{array} \right ) \otimes | \downarrow \rangle $ (introducing the pseudospin as a column vector), exhibiting a spin-pseudospin locking effect (see discussion associated to Figs. S1 and S2 in the Supplementary Information), and therefore entangled spin and pseudospin dynamics.

The reason for overlooking the role of the pseudospin on the spin dynamics is perhaps rooted in the fact that the spin transport properties appear remarkably similar to those found in common metals and semiconductors \cite{fabian2004}. Indeed, spin precession measurements in nonlocal devices result in experimental signatures that would be indistinguishable from those obtained in a metal such as aluminium \cite{jedema2002}, or a semiconductor such as GaAs \cite{crowell2007}, with extracted spin relaxation times $\tau_{s}$ that are also typically of the same order of magnitude (a few nanoseconds or lower). 
Spin relaxation in graphene has therefore been interpreted using the conventional experimental manifestations of either the Elliot-Yafet (EY) or Dyakonov-Perel (DP) mechanism \cite{Avsar2011,Han2011,Zomer2012,Dlubak2012,Ochoa2012,Zhang2012}. 
In the EY scenario, the spin relaxation time is determined by the spin mixing of carriers and the SOC of the scattering potential, and thus it is usually assumed to be proportional to the momentum relaxation time as $\tau_s \approx \alpha \cdot \tau_p$ , with $\alpha\gg 1$ (in alkali metals $\alpha\sim 10^{4}-10^{6}$) \cite{fabian2004}. 
In contrast, in the DP mechanism spin precesses about an effective magnetic field whose orientation is fixed by the momentum direction during free propagation of electrons. Such orientation changes at each scattering event, which results in a different scaling behavior as $1/\tau_{s}^{DP}\sim \Omega^{2}\tau_{p}$ \cite{fabian2004} (with $\Omega$ the average magnitude of the intrinsic Larmor frequency over the momentum distribution). Experimental estimates of $\tau_{s}$ and $\tau_{p}$ are generally obtained in a phenomenological way by fitting the experimental resistivity curves to the theoretical formulae obtained using semi-classical transport equations \cite{Tombros2007,Han2011}. However, this phenomenological analysis is not well connected with the microscopic interpretation. First of all, the weak SOC in graphene would suggest $\tau_{s}$ in the microsecond range \cite{Min2006,Ertler2009,CastroNeto2009b}, in clear disagreement with experimental data. In addition, the $\tau_{s}$ estimated in high-mobility graphene with long mean free paths remains unsatisfactorily interpreted with a single relaxation mechanism, say EY or DP \cite{Zomer2012,diniz12,Neumann13}. The suppression of $\tau_{s}$ in clean graphene has been tentatively associated to an enhanced (intrinsic or extrinsic) spin-orbit coupling due to mechanical deformations such as ripples \cite{Dery2013}, or unavoidable adatoms incorporated during the device fabrication process \cite{Pi2010, Kochan2013}, but the ultimate and microscopic nature of spin relaxation at play remains controversial and elusive \cite{Roche2014}.

Here, we explore this key fundamental issue by investigating the effect of weak perturbation induced by low densities of ad-atoms (down to $10^{12}{\rm cm}^{-2}$), which introduce a random Rashba field in real space but vanishingly small intervalley scattering, yielding long mean free paths. For typical electron densities within $[10^{10}, 10^{12}]{\rm cm}^{-2}$, the Fermi wavelength ($\lambda_{F}=2\sqrt{\pi/n}$, $n$ the charge density) lies between 20 and 200 nm and thus exceeds the mean separation between adatoms ($\sim 10$nm) where spin-orbit scattering occurs, which questions the use of a standard semiclassical description. To study spin dynamics (and spin relaxation), we use a non-perturbative method by solving the full time-dependent evolution of initially spin polarized wavepackets, either through a direct diagonalization of a continuum model, or a real space algorithm and a tight-binding model for a microscopic disorder. We describe the system of a graphene monolayer functionalized with a random distribution of adatoms. The electronic structure of clean graphene is captured by the usual $\pi$-$\pi$* orthogonal tight-binding model (with a single $p_z$-orbital per carbon site, zero onsite energies and nearest neighbors hopping $\gamma_0$). The presence of non-magnetic adatoms randomly adsorbed at the hollow positions on the graphene sheet introduces additional local spin-orbit coupling terms (Fig. \ref{Fig1}a,b), defined as \cite{Weeks2011}.

 \begin{eqnarray}
{\mathcal{H}}=&-&\gamma_0\sum_{\langle ij\rangle }c_i^+c_j+\frac{2i}{\sqrt{3}}V_{I}\sum_{\langle\langle ij\rangle\rangle \in \mathcal{R}}c_i^+\vec{s}\cdot(\vec{d}_{kj}\times\vec{d}_{ik})c_j\nonumber\\
&+&iV_R\sum_{\langle ij\rangle \in \mathcal{R}}c_i^+\vec{z}\cdot(\vec{s}\times\vec{d}_{ij})c_j-\mu\sum_{i\in \mathcal{R}}c_i^+c_i
\label{Hamil}
\end{eqnarray}

The first term is the nearest neighbor hopping with $\gamma_0 = 2.7$ eV. The second term is a complex next nearest neighbor hopping term which represents the intrinsic SOC induced by adatoms, with $\vec{d}_{kj}$ and $\vec{d_{ik}}$ the unit vectors along the two bonds connecting second neighbors, $\vec{s}$ is a vector defined by the Pauli matrices
$(s_{x},s_{y},s_{z})$, and $V_{I}$ the intrinsic SOC strength. The third term describes the Rashba SOC  ($V_R$) which explicitly violates $\vec{z} \rightarrow -\vec{z}$ symmetry, with $\vec{z}$ being a unit vector normal to the graphene plane.  The last term denotes a potential shift $\mu$ associated with the carbon atoms in the random plaquettes $\mathcal{R}$ adjacent to adatoms (Fig. \ref{Fig1}b). Such shift is due to weak electrostatic effects that arise from charge redistribution induced very locally around the adatom \cite{Weeks2011}. 

A Rashba splitting has been observed experimentally at the graphene/nickel and graphene/gold (Au) interfaces with spin splitting of up to 100 meV \cite{Dedkov2008, Marchenko2012}. Gold and nickel as well as other materials like titanium, cobalt or chromium, are usually present during the fabrication of the nonlocal spin valves that are used to determine $\tau_s$ and likely leave residues on the exposed graphene surface. Hereafter, we consider the case of Au adatoms whose influence on the transport properties of graphene has been studied experimentally \cite{Pi2010}. The tight-binding parameters to describe both intrinsic and Rashba SOC induced by such adatoms are extracted from {\it ab-initio} calculations \cite{Marchenko2012}. We then explore how the spin relaxation times scale as a function of the adatom density and adatom-induced local potential shift.

The spin dynamics is investigated by computing the time-dependence of the spin polarization defined by 
\begin{equation}
{\vec{S}}(E,t)=\frac{\langle\Psi(t)\rvert \vec{s}\delta(E-{\mathcal{H}})+\delta(E-{\mathcal{H}})\vec{s}~\rvert\Psi(t)\rangle}{2\langle\Psi(t)\rvert\delta(E-{\mathcal{H}})\rvert\Psi(t)\rangle}
\label{time-dependence0}
\end{equation}
assuming that spins are initially injected out-of-plane ($z$ direction), i.e.$\rvert\Psi(t=0)\rangle=\rvert\psi_{\uparrow}\rangle$. The time evolution of the wavepackets $\rvert\Psi(t)\rangle$ is obtained by solving the time-dependent Schr\"{o}dinger equation and the diffusion coefficients $D_x(E,t) =\frac{d}{dt}\Delta X^{ 2}(E,t)$ are evaluated from the spreading of wavepackets by using real space propagation methods \cite{Roche2012}. We focus on the expectation value of the spin z-component $S_{z}(E,t)$.  Figure \ref{Fig1} shows the typical behavior of $S_{z}(E,t)$ for two selected energies (at the Dirac point and at $E=150$ meV) and two adatom densities $\rho=0.05\%$ (about $10^{12}$ adatoms per ${\rm cm}^{2}$) (c) and $\rho=8\%$ (d). The time dependence of $S_{z}(E,t)$ is very well described by $\cos(2\pi t/T_{\Omega})e^{-t/\tau_s}$, introducing the spin precession period $T_{\Omega}$ and the spin relaxation time $\tau_s$, which are extracted from fitting the numerical simulations (solid lines). The time dependence of the modulus of the full spin polarization vector $|\vec{S}| = |(\langle s_x \rangle,\langle s_y \rangle, \langle s_z \rangle)|$ also exhibits an unambiguous signature of spin relaxation (Supplementary Fig. S3). 
Figure \ref{Fig2} gives $\tau_{s}$ and $T_{\Omega}$ extracted from the fits of $S_{z}(E,t)$ for varying adatom density. One first observes that the spin precession period is energy independent and is precisely equal to $T_{\Omega}=\pi\hbar/\bar{\lambda}_{R}$ (with $\bar{\lambda}_{R}=3\rho V_{R}$ an average SOC strength) even for the lowest coverage, which agrees with the estimate based on the continuum model \cite{Ertler2009}. In contrast, the spin relaxation time displays a significant energy dependence. A V-shape is obtained for low energy, with  $\tau_{s}$ being minimal at the Dirac point with values ranging from 0.1 ps to 200 ps when tuning the adatom density from $8\%$ to $0.05\%$ (as given in Fig.\ref{Fig3}a, main frame).  Based on the observed scaling  $\tau_{s}\sim 1/\rho$ (see Fig. \ref{Fig3}b), one can further extrapolate the spin relaxation times for even smaller defect density, obtaining $\tau_s\sim 1-10 {\rm ns} $ for adsorbate densities decreasing from $10^{11}{\rm cm}^{-2}$ down to $10^{10}{\rm cm}^{-2}$. The obtained V-shaped energy dependence and the absolute values of $\tau_s$ are remarkably similar to those reported experimentally \cite{Tombros2007,Avsar2011,Han2011,Pi2010}.

The faster relaxation at the Dirac point is actually evident in Figs. 1c and 1 d. The reason for this behaviour is the decay of the coupling between the pseudospin and momentum and the enhanced contribution of the SOC interaction, which leads to spin-pseudospin entanglement. The details of the entanglement are further described in Eq. (3) below and the Supplementary Information.

As discussed above, the usual approach to discriminate between conventional Elliot-Yafet and Dyakonov-Perel relaxation mechanisms in metals and semiconductors is to scrutinize the scaling of $\tau_s$ versus $\tau_p$. Such procedure does not necessarily apply if the dominant processes that lead to momentum and scattering relaxation are not the same. For instance, in 2D membranes that respect mirror inversion symmetry, it was demonstrated that the carrier scattering by flexural phonons leads to fast spin flips but not to momentum scattering and, therefore, the spin transport is decoupled from the carrier mobility \cite{Dery2013}. In the following discussion, we show that simple EY or DP scaling is also not suitable to describe our findings.

Within our microscopic calculations, we analyze the time-dependence of the diffusion coefficient for varying energies and ad-atom densities  (Fig. \ref{Fig2}c,d). For the lowest impurity density ($0.05\%$, Fig. \ref{Fig2}c), regardless of the considered energy, $D(E,t)$ is seen to increase in time with no sign of saturation within our computational capability, indicating a ballistic-like regime for the considered timescales. Only for the largest ad-atom density ($8\%$) does $D(t)$ eventually saturate at high enough energies (above 100 meV, $D(t)\to D_\text{max}$), allowing for the evaluation of the transport time using $\tau_{p}(E)=D_\text{max}(E)/2v^{2}(E)$ (see dashed lines in Fig. \ref{Fig2}b). A sharp increase of $\tau_{p}$ is seen when approaching the Dirac point, where $\tau_s$ reaches its minimum value, with $\tau_s\ll \tau_{p}$. This energy dependence in $\tau_{p}$ is not unique to gold adatoms but has also been observed for other types of disorder with a weak intervalley scattering contribution, such as epoxide defects or long range scatterers \cite{Roche2012}. As seen in Fig. \ref{Fig3}b, $\tau_{s}\sim 1/\rho$, which does not allow us to discriminate between EY and DP processes. However, the absolute values of $\tau_s$ and $\tau_{p}$ (with $\tau_s\ll\tau_p$) are a clear manifestation of the breakdown of the typical scaling associated to both mechanisms. Even the unconventional DP regime described in Ref.\cite{fabian2004} for the case of $\tau_{p}/T_{\Omega}\geq 1$ where $1/\tau_s\sim \Delta\Omega$ (with $\Delta\Omega$ an effective width of the distribution of precession frequencies) cannot account for the observation that a weak variation in the local disorder affects the absolute values of $\tau_s$ (while $\rho$ is unchanged) as observed in Fig. \ref{Fig2}.  Here local disorder is monitored by the $\mu$ parameter. (Although $\mu$ belongs to the TB parameterization of the adatom, we use it temporarily to increase local disorder.) In fact, its value could slightly change when modifying the substrate screening or in presence of a more strongly bonded adsorbant than Au. As a consequence of the above findings, the spin relaxation mechanism at play is incompatible with both the EY and the DP mechanisms, a fact which could shed new light on the current debate on the microscopic nature of spin relaxation in clean graphene \cite{Zomer2012,diniz12,Neumann13}.

We now further study the origin of the $\tau_{s}$ minimum at low energy and its unconventional scaling with $\tau_p$. Given that our simulations with the microscopic model give $\tau_{s}\ll\tau_{p}$, we further explore the low-energy spin dynamics with an effective continuum model, in which the spin-orbit scattering is treated as a homogeneous potential \cite{Ertler2009}. We solve the Dirac equation in the continuum model by using a 4$\times$ 4 effective Hamiltonian, taking into account the pseudospin degree of freedom 

\begin{equation}
 h(\vec{k})=h_0(\vec{k})+h_R(\vec{k})+h_I(\vec{k})
\label{SOCHal}
\end{equation}
While the hopping from three nearest neighbors $h_0(\vec{k})= \hbar v_F(\zeta \sigma_xk_x+\sigma_yk_y) \otimes 1_s$ dominates at high energy and vanishes at the Dirac point ($\zeta=\pm 1$ for $K$ and $K'$ valleys, $\vec{\sigma}$ are pseudospin Pauli matrices and $1_s$ is a $2 \times 2$ identity matrix), the intrinsic SOC $h_I(\vec{k})= \bar{\lambda}_I \zeta\left[ \sigma_z \otimes s_z \right]$ and the Rashba interaction $h_R(\vec{k})=  \bar{\lambda}_R \left( \zeta \left[ \sigma_x \otimes s_y \right] - \left[ \sigma_y \otimes s_x \right] \right)$ play an extremely important role at the Dirac point, where the coupling between spin and pseudospin becomes predominant, and governs the quantum dynamics and dephasing of the wavepackets as described below. 

Within the continuum model spin relaxation is achieved by introducing an \textit{ad-hoc} energy broadening. We use an initially $z$-polarized state for injection and consider only the $K$ valley. A certain density of Au impurities (inducing local spin-orbit coupling) is described by the effective spin-orbit coupling $\bar{\lambda}_{R}=3\rho V_R$  and $\bar{\lambda}_I=3\sqrt{3}\rho V_I$. Note that no additional local (static) scattering potential is introduced here ($\mu = 0$). By computing the spin dynamics of initially spin-polarized wavepackets, one also obtains a spin relaxation effect defined by the two timescales $T_\Omega$ and $\tau_{s}$ (see Supplementary Material).

It is instructive to contrast the results of the continuum model (Fig. \ref{Fig3}a, inset) with those from the microscopic model (Fig. \ref{Fig3}a, main frame). Although the spin precession period $T_\Omega$ obtained by both models is identical (Fig. \ref{Fig3}b) and the energy dependence of $\tau_{s}$ is similar, the absolute values of $\tau_{s}$ differ substantially, especially in the high energy regime, where $\tau_{s}$ is clearly overestimated using the continuum model. Such difference also becomes increasingly large upon decreasing the adatom density because $\tau_{s}$ presents a different scaling with defect coverage (Fig.\ref{Fig3}b). This clearly evidences the importance of disorder and illustrates the limits of a phenomenological approach using the continuum model for a quantitative comparison with experimental data. Notwithstanding, the qualitative agreement between both models (particularly for high coverage) and the weak momentum relaxation effects observed in the microscopic model (as seen in the long $\tau_p$) suggest some generality in the unconventional spin relaxation observed near the Dirac point.

To further substantiate the origin of the spin relaxation, we scrutinize the spin and pseudospin dynamics of wavepackets using the continuum model. Pseudospin is intrinsically related to the graphene sublattice degeneracy and, as long as valley mixing is negligible, pseudospin is aligned in the direction of the momentum at high energy ($h_0(\vec{k})$ dominates the Hamiltonian \eqref{SOCHal}). The Rashba spin-orbit term $h_R(\vec{k})$ entangles spin $\vec{s}$ with the lattice pseudospin $\vec{\sigma}$, overriding the locking rule between pseudospin and momentum since $h_0(\vec{k})$ becomes vanishingly small in the vicinity of the Dirac point \cite{Rashba2009,Ochoa2012}. 

Figure \ref{Fig4} highlights the spin dynamics at different low ($E=0$, $E=-5$ meV) and high ($E=130$ meV) energies, which are representative of the underlying physics (note that no relaxation takes place for fixed energy, thus the requirement of the \textit{ad-hoc} broadening). At high energy, the spin precesses quite regularly showing an oscillatory pattern of $S_z(t)$  dominated by a single period $T_\Omega=\pi\hbar /\overline{\lambda}_R =$0.19 ps (Fig. \ref{Fig4}a). The spin precession occurs about an effective magnetic field $B_R$ dictated by the Rashba SOC and pointing tangentially to the Fermi circle (as seen from the precession from blue to pink in middle panels from $t_{1}$ to $t_{4}$).  In contrast, the pseudospin $\langle \vec{\sigma}(t)\rangle$ points approximately in the same direction of the momentum (evolving from green to orange). Its oscillatory pattern is driven by the Rashba period $T_\Omega$ together with a superimposed and more rapid oscillation (see Supplementary Material).

The situation at low energy is markedly different (Fig. \ref{Fig4}b,c). We observe a highly unconventional spin and pseudospin motion which is analyzed more closely for the spin and pseudospin $z$-components at the Dirac point and at $E=-5$ meV. Here, the amplitude of the pseudospin oscillation is strongly enhanced since pseudospin is no longer locked with momentum but starts to precess about an effective pseudo-magnetic field. The pseudo-magnetic field strongly depends on the spin orientation, thus yielding complex time-dependent dynamics of spin and pseudospin (see middle panels of Figure \ref{Fig4} corresponding to \ref{Fig4}b,c). Such an effect derives from the increased pseudospin precession period $T^\text{ps}_0 = \pi\hbar/E$ (about $B_0^\text {ps}$), which decays significantly at low energy.
Therefore $\langle \sigma_i\rangle$ can no longer be replaced by its time average $\overline{\langle \sigma_i\rangle}$, which in consequence also holds for the Rashba field $B_R$. The time dependence of $B_R$ with variability on a timescale similar to the Rashba period leads then to strong non-linear dynamics of spin and pseudospin motion.
As a result of such coupled dynamics, the spin precession cannot be described by a single period $T_\Omega$ as becomes evident from the complex Fourier spectra of $S_z(t)$ in Fig. \ref{Fig4}d.  The time dependence of $B_R$ includes also changes of its direction, thus impacting the pseudospin and lifting the pseudospin-momentum locking. Both of these effects finally produce a joint spin/pseudospin motion prohibiting the de-coupling of driving forces ($B^\text{ ps}_0$, $B_R$) that was possible at higher energies.

While the continuum model provides qualitative insight into the spin-pseudospin coupling and entanglement of their corresponding wavefunctions, the microscopic model enables the quantification of spin relaxation times for a given microscopic disorder. By scrutinizing the general form of the spin polarization (Eq. \ref{time-dependence0}), a simple understanding of the spin relaxation mechanism can be drawn. In the microscopic model, the propagation of an initially spin-polarized wavepacket  $\rvert\psi_{\uparrow}(t=0)\rangle$, is driven by the evolution operator $e^{-i{\mathcal{H}}t/\hbar}\rvert\psi_{\uparrow}(t=0)\rangle$, with ${\mathcal{H}}$ consisting of the clean graphene term plus the SOC potential, which acts as a local (and random) perturbation on the electron spin. The time-dependence of the total spin polarization results from the accumulated dephasing along scattering trajectories developed under the evolution operator. As the distribution of scattering centers is random in space, all different trajectories accumulate different phase shifts in their wavefunctions (each being the result of local spin/pseudospin coupling and disorder potential). When phase shifts for up and down components average out, the spin polarization of $\rvert\psi_{\uparrow}(t=0)\rangle$ is lost.

In conclusion, our spin transport study in chemically modified graphene has revealed a hitherto unknown phenomenon related to the entangled dynamics of spin and pseudospin, induced by spin-orbit coupling and leading to fast spin relaxation in a quasi-ballistic transport regime. Entanglement between spin and orbital degrees of freedom has been discussed for ballistic semiconducting nanowires \cite{Nikolic2005}. Here, faster spin relaxation develops when spin-pseudospin entanglement is maximized at the Dirac point, where the momentum scattering time becomes increasingly large because disorder preserves pseudospin symmetry. Such mechanism, occurring in clean graphene with long mean free paths, has no equivalent in condensed matter and cannot be described by EY or DP. It is here described for gold adsorbates, but should also be at play for other sources of local spin-orbit coupling (ripples, defects, etc.), thus contributing to the understanding of spin transport in graphene devices \cite{Tombros2007,Avsar2011,Han2011,Zomer2012, Dlubak2012}. Finally, such finding could open the path to control the spin by modifying the pseudospin or vice versa. For example, spins could be manipulated by inducing pseudomagnetic fields by straining graphene. Such possibilities suggest unprecedented approaches for the emergence of non-charge-based information processing and computing, resulting in a new generation of active (CMOS-compatible) spintronic devices together with non-volatile low-energy MRAM memories \cite{Dery2012}.

\section{Methods}
\noindent
{\bf Derivation of Eq.(\ref{time-dependence0})}. The time dependence of the spin polarization of states in graphene $\pi_z$ bands can be derived from the  expectation value of the Pauli spin operator
\begin{equation}
\langle  {\vec {s}}(t)\rangle={\rm Tr}\left[\rho(0) {\vec{s}}(t)\right]
\label{Eq:spin_band}
\end{equation}
where the density matrix $\rho(0)$ accounts for the initial spin polarization (out-of-plane) and 
$ {\vec {s}}(t)=e^{\frac{i {\mathcal{H}} t}{\hbar}}  {\vec {s}} e^{\frac{-i { \mathcal{H}} t}{\hbar}}$ is the spin operator in Heisenberg representation.
As any trace, it can be replaced by the expectation value with random-phase states according to ${\rm Tr}\left[\cdots\right]\to\langle\varphi_\text{RP}'\lvert\cdots\rvert\varphi_\text{RP}'\rangle$ where the random-phase state
$|\varphi_\text{RP}'\rangle=
\frac{1}{\sqrt{2M}}\sum_{j=1}^M  \left(\begin{array}{c}e^{i\theta_j}\\
 e^{i\theta_j'}\end{array}\right) \rvert j\rangle$
is not an energy eigenstate at Fermi energy but samples the full spectrum. The trace in Eq. \eqref{Eq:spin_band} (and equally the form with $\lvert\varphi_\text{RP}'\rangle$) includes all states of the system at higher and lower energies, and not only those accessible in transport experiments (which are restricted to the Fermi energy). Accordingly, Eq. \eqref{Eq:spin_band} is not appropriate when aiming at a comparison to experiment and another quantity needs to be computed as explained below. The quantum average of a given operator $ {Q}$ at a selected energy $E$ can be generally written as average over all eigenstates at this energy through
\begin{equation}
 \langle  {Q}\rangle_E=\frac{1}{N}\sum_i^N \langle \psi_E^i\lvert  {Q} \rvert \psi_E^i\rangle
=\frac{{\rm Tr}\left[\sum_i^N  \rvert \psi_E^i\rangle\langle \psi_E^i\lvert  {Q}\right]}{N}
 \label{QuanAverage}
\end{equation}
where $\rvert \psi_E^i\rangle$ are $N$ degenerate eigenstates of $ {\mathcal{H}}$ at energy $E$ which are obtained by Hamiltonian diagonalization.  It is next straightforward to show that
\begin{equation}
 \langle  {Q}\rangle_E=\frac{1}{N}\sum_i^N \langle \psi_E^i\lvert  {Q} \rvert \psi_E^i\rangle
=\frac{{\rm Tr}\left[\delta(E- {\mathcal{H}}) {Q}\right]}{{\rm Tr}\left[\delta(E- {\mathcal{H}})\right]}
=\frac{{\rm Tr}\left[\delta(E- {\mathcal{H}}) {Q}+ {Q}\delta(E- {\mathcal{H}})\right]}{2{\rm Tr}\left[\delta(E- {\mathcal{H}})\right]},
 \label{QuanAverage}
\end{equation}
where $\delta(E- {\mathcal{H}})$ is the continous projection operator, and the normalization with the number of states $N$ at energy $E$ is replaced by the density of states ${\rm Tr}\left[\delta(E- {\mathcal{H}})\right]$ at this energy. The last equality in Eq.(\ref{QuanAverage}) yields a 
symmetric (Hermitian) form in the numerator suitable when $ {Q}$ does not commute with the Hamiltonian. While the case of an average over unpolarized states in Eq.(\ref{QuanAverage}) yields
\begin{equation}
\begin{aligned}
\langle  {\vec{s}}(t)\rangle_E&=\frac{{\rm Tr}\left[\delta(E- {\mathcal{H}}) {\vec {s}}(t)\right]}{{\rm Tr}\left[\delta(E- {\mathcal{H}})\right]}
=\frac{{\rm Tr}\left[\delta(E- {\mathcal{H}}) {\vec {s}}(t)+ {\vec {s}}(t)\delta(E- {\mathcal{H}})\right]}{2{\rm Tr}\left[\delta(E- {\mathcal{H}})\right]}\\
&=\frac{\langle\varphi_\text{RP}'\lvert \delta(E- {\mathcal{H}}) {\vec {s}}(t)+ {\vec {s}}(t)\delta(E- {\mathcal{H}}) \rvert\varphi_\text{RP}'\rangle}
{2\langle\varphi_\text{RP}'\lvert \delta(E- {\mathcal{H}})\rvert\varphi_\text{RP}'\rangle},
\end{aligned}
\label{s}
\end{equation}
the trace with spin-polarized initial random phase states $\rvert\varphi_{RP}\rangle=\frac{1}{\sqrt{N}}\sum_{j=1}^N 
\left(\begin{array}{c}e^{i\theta_j}\\
 0\end{array}\right)\rvert j\rangle$, which is of interest here, yields

\begin{equation}
{\vec{S}}(E,t)=\frac{\langle\Psi(t)\rvert \vec{s}\delta(E- {\mathcal{H}})+\delta(E- {\mathcal{H}})\vec{s}~\rvert\Psi(t)\rangle}{2\langle\Psi(t)\rvert\delta(E- {\mathcal{H}})\rvert\Psi(t)\rangle}
\end{equation}
where the time evolution of the wavepackets $\rvert\Psi(t)\rangle=e^{\frac{-i {\mathcal{H}} t}{\hbar}}\rvert\Psi(0)\rangle\equiv e^{\frac{-i { \mathcal{H}} t}{\hbar}}\rvert \varphi_{RP} \rangle$ is obtained by solving the time-dependent Schr\"{o}dinger equation.
\\
\noindent
{\bf Real space implementation of the wavepacket quantum dynamics}. The transport times are deduced from the numerical analysis of the spreading of wavepacket through \cite{Roche2012,FoaTorres2014}:

\begin{equation}
\Delta X^{2}(E,t) = \frac{\displaystyle {\large\rm Tr}\bigl[ \delta(E- {\mathcal{H}}) |  {X}(t)-  {X}(0) |^2 \bigr]}
{ \strut\displaystyle {\large\rm Tr}[\delta(E- {\mathcal{H}})]}
\label{DeltaX2}
\end{equation}

A key quantity in the analysis of the transport properties is the diffusion coefficient: $D_x(E,t) =\frac{d}{dt}\Delta X^{ 2}(E_{F},t)$. Assuming the system to be isotropic for the in-plane $x$ and $y$ directions, then $D(t) = D_x(t) + D_y(t) = 2D_x(t).$ The diffusion coefficient contains all information about the semiclassical effects of scattering leading to diffusive behavior, but also the quantum interference effects which lead to localization effects. $D(t)$ increases ballistically at short times, then saturates due to elastic scattering events, and finally decays as a result of quantum interference effects (when significant). The elastic mean free path is derived from the maximum of the diffusion coefficient:  $\ell_e(E)=D^{\text{max}}(E)/2 v(E)$ , with $v(E)$ being the carrier velocity and $D^{\text{max}}$ the maximum value of $D(t)$. The momentum relaxation can be extracted from elastic mean free path $\tau_p(E)=\ell_e(E)/v(E)$.

\section{Acknowledgments}
We thank Aron Cummings for a critical reading of the manuscript. The research leading to these results has received funding from the European Union Seventh Framework Programme under grant agreement number 604391 Graphene Flagship. This work was also funded by Spanish Ministry of Economy and Competitiveness under contract MAT2012-33911 and MAT2010-18065. S.O.V. acknowledges ERC Grant agreement 308023 SPINBOUND.

\section{Author contributions}
D.V.T., D.S. and F.O. designed the models and performed the calculations.  D.V.T., F.O., S.O.V. and S.R. carried out analyses and interpretation. F.O., S.O.V. and S.R. wrote the text and all authors contributed to the manuscript and supplementary information.

\section{Additional information}
The authors declare no competing financial interests.  Reprints and
permissions information available online at http://npg.nature.com/reprintsandpermissions.
Correspondence and requests for materials should be addressed to S.R.

\begin{figure}[htbp]
\begin{center}
\leavevmode
\caption{{\bf Spin Dynamics in disordered graphene}. (a) Ball-and-stick model of a random distribution of adatoms on top of graphene (b) Top view of the gold adatom sitting on the center of an hexagon  (c),(d) Time-dependent projected spin polarization $S_{z}(E,t)$ of charge carriers (symbols) initially prepared in an out-of-plane polarization  (at Dirac point (red curves) and at $E=150$ meV (blue curves)). Analytical fits are given as solid lines (see text).  Parameters are $V_I=0.007\gamma_0$, $V_R=0.0165\gamma_0$, $\mu=0.1\gamma_0$, $\rho=0.05\%$ (c) and $\rho=8\%$ (d).}
\label{Fig1}
\end{center}
\end{figure}

\begin{figure}[htbp]
\begin{center}
\caption{{\bf Spin relaxation times and transport mechanisms}. Spin relaxation times ($\tau_s$) for $\rho=0.05\%$ (a) and $\rho=8\%$ (b). Black (red) solid symbols indicate $\tau_s$ for $\mu=0.1\gamma_0$ ($\mu=0.2\gamma_0$).
$T_{\Omega}$ vs. $E$ is also shown (open symbols). $\tau_p$ (dotted line in (b)) is shown over a wider energy range (top $x$-axis) to stress the divergence around $E=0$ ($\mu=0.2\gamma_0$). Panels (c) and (d): Time dependent diffusion coefficient $D(t)$ for $\rho=0.05\%$ and $\rho=8\%$ with $\mu=0.2\gamma_0$.  }
\label{Fig2}
\end{center}
\end{figure}

\begin{figure}[htbp]
\begin{center}
\caption{{\bf Spin relaxation times deduced from the continuum and microscopic models}. (a) Spin relaxation times ($\tau_s$) for varying $\rho$ between 0.05$\%$ and 8$\%$ extracted from the microscopic model (with $\mu=0.1\gamma_0$). Inset: $\tau_s$ values using the continuum model for $\rho=1\%$ and 8$\%$ (filled symbols). A comparison with the microscopic model (with $\mu=0$) is also given for $\rho=8\%$ (open circles). (b) Scaling behavior of $T_{\Omega}$ and $\tau_s$ versus $1/\rho$. The $T_{\Omega}$ values obtained with the microscopic (resp. continuum) model are given by red diamonds (resp. red solid lines).  $\tau_s$ values for the microscopic model (blue squares) and the continuum model (black circles) are shown for two selected energies $E=150$meV (solid symbols) and $E=0$ (open symbols). Solid lines are here guides to the eye.}
\label{Fig3}
\end{center}
\end{figure}

\begin{figure}[htbp]
\begin{center}
\caption{{\bf Spin and pseudospin dynamics in graphene with $\rho=8\%$ of adatoms}.
Time dependence of spin-polarization $S_z$ (blue) and pseudospin polarization $\sigma_z$ (green) in $z$ projection for energies $E=130$meV (a), $E=0$ (b), and $E=-5$ meV (c). Note that all quantities are normalized to their maximum value to better contrast them in the same scale. Middle panels show the time evolution for both spin (from blue to pink) and pseudospin (from green to orange). The snapshots are taken at different times from $t_{1}$ to $t_{4}$ sampling the shaded regions in (a), (b), (c). (d) Fourier transform of $S_z(t)$ plotted over oscillation period, and showing non-dispersive spectra at high energy (between $E=$125 meV, 130 meV and 135 meV). Low-energy spectra (for $E=-5$ meV, 0 and 5 meV) change strongly with energy (dispersive) showing a gradual reduction and blue shift of the original Rashba peak at about 0.19 ps and the appearance of additional features. }
\label{Fig4}
\end{center}
\end{figure}

\end{document}